# Observation of inhomogeneous domain nucleation in epitaxial Pb(Zr,Ti)O$_3$ capacitors

D. J. Kim[a)], J. Y. Jo, T. H. Kim, S. M. Yang, B. Chen, Y. S. Kim, and T. W. Noh[b)]

*ReCOE & FPRD, Department of Physics and Astronomy, Seoul National University,*

*Seoul 151-747, Korea*

We investigated domain nucleation process in epitaxial Pb(Zr,Ti)O$_3$ capacitors under a modified piezoresponse force microscope. We obtained domain evolution images during polarization switching process and observed that domain nucleation occurs at particular sites. This inhomogeneous nucleation process should play an important role in an early stage of switching and under a high electric field. We found that the number of nuclei is linearly proportional to log(switching time), suggesting a broad distribution of activation energies for nucleation. The nucleation sites for a positive bias differ from those for a negative bias, indicating that most nucleation sites are located at ferroelectric/electrode interfaces.

PACS: 77.84.–s, 77.80.Fm, 77.65.–j, 68.37.–d

---

[a)] Present address: ULP/UMR CNRS 7504, IPCMS – GEMME, 23 rue du Loess, 67034 STRASBOURG Cedex 2, France; electronic mail: dong-jik.kim@ipcms.u-strasbg.fr
[b)] Electronic mail: twnoh@snu.ac.kr

When applying an electric field ($E$) to a ferroelectric material, polarization ($P$) switching occurs via the nucleation of domains with reversed $P$ and subsequent domain wall motion.[1,2] Domain nucleation is the first necessary step for $P$ switching, but creating a nucleus randomly inside a ferroelectric material is quite difficult. In the 1950s, Landauer pointed out that the thermodynamic energy barrier ($U^{th}$) for creating such a nucleus should be practically insurmountable: $U^{th} > 10^3 \, k_B T$ under $E \approx 100$ kV/cm. To overcome what is known as ``Landauer's paradox,'' many studies assumed that the nucleus creation of reversed $P$ occurs nonrandomly at particular sites, probably due to defects.[3-7] This concept of inhomogeneous nucleation has been widely accepted[3,4] and dealt with in numerous theoretical models,[5-7] but experimental data in support of this idea are lacking.

To investigate the inhomogeneous nucleation process in epitaxial PbZr$_{0.4}$Ti$_{0.6}$O$_3$ (PZT) films visually, we used the step-by-step switching approach under piezoresponse force microscopy (PFM) suggested by Gruverman *et al.*[8,9] The main improvement in our setup was that we adopted a separate probe needle to obtain reliable electric contact with the top electrode, as depicted in Fig. 1(a). Since the PFM cantilever only detects piezoelectric vibration, we could obtain reliable domain images of a large area in ferroelectric capacitors.

To fabricate our PZT capacitors, we deposited SrRuO$_3$/PZT/SrRuO$_3$ epitaxial layers *in-situ* on SrTiO$_3$ (001) substrates using a pulsed laser deposition method. The PZT and SrRuO$_3$ top electrode layers were 170 and 20 nm thick, respectively. AFM images showed very clean surface after the deposition of each layer and HRXRD results confirmed the epitaxiy of all layers in our sample. The top electrode layer was necessary for applying a uniform $E$ to the PZT film, but its thickness determined the spatial resolution of the resulting PFM images. With a 20-nm-thick top electrode layer, we obtained a spatial resolution of about 10 nm. We fabricated PZT capacitors with an area of $100 \times 100$ μm$^2$ using reactive ion etching. The inset of Fig. 1(c) shows that the PZT capacitor resulted in a good *P-E* hysteresis loop with a remnant polarization ($P_r$) of 49 μC/cm$^2$.

With the experimental setup shown in Fig. 1(a), we could measure the value of the switched *P* and associated domain images. We used an arbitrary waveform generator (FG300, Yokogawa) to apply square pulses to the top electrode via the probe needle. Figure 1(b) shows a typical pulse sequence used for the step-by-step switching.[2,3,8] We used the resetting pulses to pole the PZT capacitor fully in one direction. We increased the duration ($t_{sw}$) of the switching pulses incrementally, so that we could increase the amount of reversely polarized volume systematically. To measure the amount of change in *P* (Δ*P*), we read the switching current responses using a digital oscilloscope (DL7100, Yokogawa).[2] After each switching pulse, we scanned $6 \times 6$ μm$^2$ on the top SrRuO$_3$ electrode using the PFM (XE-100, Park Systems). To obtain piezoelectric responses, we applied an ac-field modulation (0.2 V$_{rms}$, 19.1 kHz) to the top electrode via the probe needle and confirmed that the 0.2 V$_{rms}$ ac-field did not affect the domain configurations in our 170-nm-thick epitaxial PZT capacitors.[8] We measured the amplitude (*R*) and phase (*θ*) of the piezoelectric signals using a lock-in amplifier (SR830, Stanford Research Systems). We used the WSxM program to analyze all of the domain images from the PFM studies.[10]

We could determine the volume fraction (*q*) of the reversed domains from the piezoelectric signal. In our setup, the amount of the reversed domains near a particular position should be proportional to *R*cos*θ*. We obtained *q* by summing the *R*cos*θ* signals over the scan area and normalizing it properly, since $R\cos\theta \propto d_{33} = 2Q_{33}\varepsilon\varepsilon_0 P$, where $d_{33}$ is the piezoelectric constant, $Q_{33}$ is the electrostriction coefficient, and $\varepsilon$ and $\varepsilon_0$ are the dielectric constant of the ferroelectric layer and dielectric permittivity of a vacuum, respectively. The symbols in Fig. 1(c) denote the obtained $q$ values at various values of $t_{sw}$ and $E$. Note that these $q$ values agreed well with values of $\Delta p$ ($\equiv \Delta P/2P_r$) from the switching current measurements. This excellent agreement demonstrates the validity of our experimental technique.

Our PFM studies showed that the reversed domains started to merge when $\Delta p$ was about $0.3 \pm 0.1$. To investigate the nucleation process, we looked at the early stage of $P$ switching, namely when $\Delta p$ was about 0.1. We obtained 30 phase images of a $6 \times 6$ μm² area by scanning the same area after resetting and switching with $E = 90$ kV/cm and $t_{sw} = 1.5$ μs each time. Figure 2(a) shows the spatial distribution of the nucleation probabilities obtained by summing all of the phase images. Darker spots indicate that nucleation occurred more frequently at that position. The white area indicates that no $P$ reversal occurred there during the 30 repeats. Figure 2(b) shows the number of sites for a given probability in the 30 repeated measurements. Most nucleation sites had a probability $> 0.9$, at which nuclei appeared more than 27 times in 30 repeats. If the probability of a nucleation site was 0.9 with $E = 90$ kV/cm and $t_{sw} = 1.5$ μs, the activation energy of that site should have been about 10 $k_B T$, which was much less than $U^{th}$ of about $10^3$ $k_B T$. Our PFM studies clearly demonstrated that nucleation should occur inhomogeneously in our epitaxial PZT capacitor.

In the $P$ switching process, the role of nucleation became more important at the earlier stage and at higher $E$. Figure 2(c) shows the number of nuclei and $\Delta p$ at various $t_{sw}$ with $E = 60$ kV/cm. When $t_{sw} < 10^{-5}$ s, $\Delta p$ was linearly proportional to the nucleus number. However, when $t_{sw} > 10^{-5}$ s, $\Delta p$ increased more rapidly, indicating that the domain wall motion started to contribute more to the $P$ switching. From Fig. 2(a), it is clear that most $P$ changes at the early stage were due to domain nucleation. Figure 2(d) shows the $E$-dependence of nucleus number at various $P$ switching stages, which are characterized by the $\Delta p$ values. The nucleus number increased with $E$, indicating that nucleation played a more important role at higher $E$.

During the early stage of $P$ switching, the nucleus number increased linearly to log $t_{sw}$, as shown in Fig. 2(e). Similar time dependences have been observed in numerous magnetic systems and explained in terms of a broad distribution of activation energies.[11] Similarly, the observed linear dependence on log $t_{sw}$ in our PZT capacitor could be interpreted in terms of a wide distribution of activation energies for nucleation. In addition, this picture is consistent with the fact that the nucleation probability had a long tail below 0.9, as shown in Fig. 2(b).

Since Landauer first proposed the idea,[5-7] it has been widely accepted that the domains nucleate at the ferroelectric/electrode interfaces. Figures 3(a) and (b) show phase images that were measured with $E = +150$ and $-150$ kV/cm, respectively. To visualize the difference in the nucleation sites with respect to the bias polarity clearly, we obtained the merged and difference images in Figs. 3(a) and (b). The merged image in Fig. 3(c) shows that most of the reversed domains under a positive $E$ nucleated at different sites from those under a negative $E$. The difference image in Fig. 3(d) shows that few sites existed in which reversed domains became nucleated under both positive and negative $E$. This suggests that nucleation is very likely to occur at the ferroelectric/electrode interfaces, rather than the inside of the film.

The inhomogeneous nucleation in our 170-nm-thick PZT capacitors seems to contradict our earlier work, which showed that homogeneous nucleation plays an important role in high-quality ultrathin BaTiO$_3$ capacitors.[12] A key to these seemingly contradictory cases is the thickness of the ferroelectric film. Inside a ferroelectric film with the capacitor geometry, the *P* charges at the ferroelectric/electrode interfaces are screened by the free carriers inside the conducting electrode. However, the screening is incomplete and induces a depolarization field. As the film becomes thinner, the depolarization field becomes larger. For a 5-nm-thick BaTiO$_3$ film, the depolarization field is about 800 kV/cm.[13] Since $U^{th}$ is proportional to $E^{-5/2}$,[6] the large depolarization field can reduce the value of $U^{th}$ to about 10 $k_BT$, so that homogeneous nucleation can occur. In contrast, in our 170-nm-thick PZT capacitors, the depolarization field is about 50 kV/cm. This value is not large enough to reduce $U^{th}$ to thermally accessible values for homogeneous nucleation, which is empirically shown by scaling behaviors of coercivity over thickness.[14] It suggests that nucleation in 170 nm thick PZT films is inhomogeneous.[15]

In summary, we investigated the nucleation behavior of 170-nm-thick epitaxial Pb(Zr,Ti)O$_3$ film using a modified piezoresponse force microscope. We obtained experimental evidence for inhomogeneous nucleation. In addition, we found that most nucleation sites were located at the ferroelectric/electrode interface and had a broad distribution of activation energies for nucleation.

We thank Prof. J.-G. Yoon, Prof. T. K. Song, and Prof. S.-B. Choe for valuable discussions. This study was supported financially by the Creative Research Initiatives (Functionally Integrated Oxide Heterostructures) of the Korean Science and Engineering Foundation (KOSEF).

[14] M. Dawber, P. Chandra, P. B. Littlewood, and J. F. Scott, J. Phys.: Condens. Matter **15**, L393 (2003).

[15] In order to connect the existing scaling behavior of coercivity with domain nucleation process, we need the barrier height. If the value of the barrier height is single-valued, we will easily derive the coercivity scaling behavior over thickness. However, in our case, we showed that barrier height has a broad distribution. In this state, we cannot state that we have a single barrier height to induce homogeneous nucleation in our PZT films.

**Figure captions**

Fig. 1 (Color online) (a) Schematic diagram of the experimental setup. (b) The pulse sequence used to study domain nucleation using the step-by-step switching scheme. To remove the imprint induced during each step, we applied 100 bipolar training pulses between the switching measurements. (c) Dependence of switching behaviors on the switching pulse width ($t_{sw}$). The lines indicate the switched $P$ fraction ($\Delta p \equiv \Delta P/2P_r$) obtained by the electrical measurements, and the symbols indicate the volume fraction ($q \propto R\cos\theta$) of the reversed domains estimated from the PFM measurements. The inset shows a typical $P$-$E$ hysteresis loop of the epitaxial $SrRuO_3$/$Pb(Zr,Ti)O_3$/$SrRuO_3$ capacitor.

Fig. 2 (Color online) (a) The spatial probability distribution of nucleation sites in a $6 \times 6$ $\mu m^2$ in epitaxial $Pb(Zr,Ti)O_3$ capacitor. (b) The numerical distribution of nucleation sites in terms of the probability obtained from (a). The dependence of nucleus numbers on (c) $t_{sw}$ and (d) $E$ in the initial stage of polarization switching. In (c), the numbers of reversed domains at $E = 60$ kV/cm are compared with $\Delta p$ at the same $E$. In (d), the numbers of reversed domains when $\Delta p$ reaches a certain value (*i.e.*, $\Delta p = 0.05, 0.10, 0.15,$ or $0.20$) are plotted. (e) The dependence of nucleus number on $t_{sw}$ at $E = 60 \sim 120$ kV/cm.

Fig. 3 (Color online) Visualization of nucleated domains from the PFM images under (a) positive ($E = +150$ kV/cm) and (b) negative ($E = -150$ kV/cm) pulses with $t_{sw} = 0.5$ $\mu s$. The nucleated domains are shown as the dark areas in (a) and as the bright areas in (b). The merged image (c) = (a) + (b) shows that most of the reversed domains under a positive pulse become nucleated at different sites from those under a negative pulse. The difference image (d) = (a) − (b) shows the sites in which reversed domains became nucleated under both positive and negative pulses.

**Fig. 1**

**D. J. Kim** *et al.*

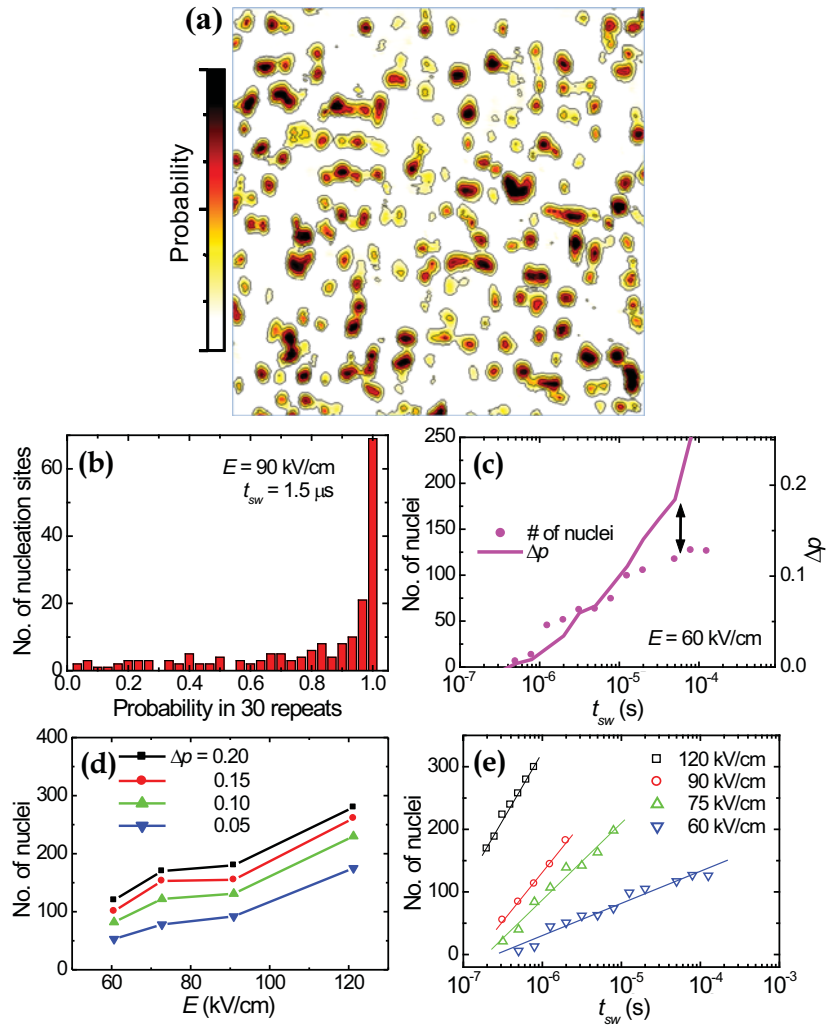

**Fig. 2**

**D. J. Kim *et al.***

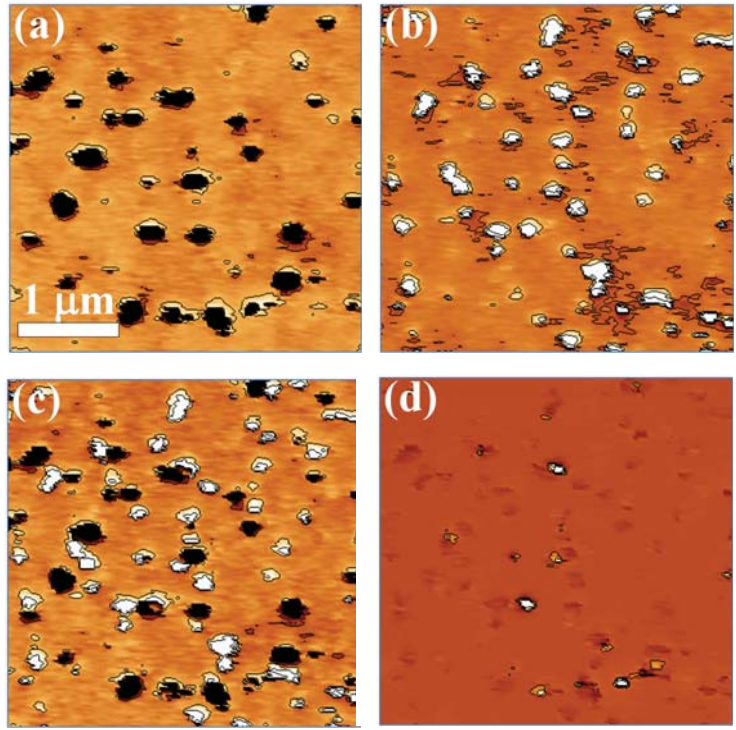

**Fig. 3**

**D. J. Kim** *et al*.